\documentclass[conference]{IEEEtran}
\IEEEoverridecommandlockouts
\usepackage{cite}
\usepackage{amsmath,amssymb,amsfonts}
\usepackage{algorithmic}
\usepackage{graphicx}
\usepackage{textcomp}
\usepackage{xcolor}
\usepackage{comment}
\usepackage{todonotes}
\usepackage{balance}
\usepackage{multirow}
\usepackage{hyperref}
\usepackage{makecell}

\def\BibTeX{{\rm B\kern-.05em{\sc i\kern-.025em b}\kern-.08em
    T\kern-.1667em\lower.7ex\hbox{E}\kern-.125emX}}
    
\begin{document}

\title{Towards Automated Generation of \\Smart Grid Cyber Range for  Cybersecurity Experiments and Training
}

\author{\IEEEauthorblockN{Daisuke Mashima, Muhammad M. Roomi, \\Bennet Ng, Zbigniew Kalbarczyk}
\IEEEauthorblockA{\textit{Illinois at Singapore Pte Ltd} \\
%Singapore\\
\{daisuke.m,roomi.s,bennet.ng,kalbarcz\}@adsc-create.edu.sg}
\and
\IEEEauthorblockN{S.M. Suhail Hussain, Ee-chien Chang}
\IEEEauthorblockA{
\textit{National Cybersecurity R\&D Lab}\\
\textit{National University of Singapore} \\
%Singapore \\
suhail@ieee.org, changec@comp.nus.edu.sg}

}

\pagestyle{plain}

\maketitle

\begin{abstract}
Assurance of cybersecurity is crucial to ensure dependability and resilience of smart power grid systems. In order to evaluate the impact of potential cyber attacks, to assess deployability and effectiveness of cybersecurity measures, and to enable hands-on exercise and training of personals, an interactive, virtual environment that emulates the behaviour of a smart grid system, namely smart grid cyber range, has been demanded by industry players as well as academia. A smart grid cyber range is typically implemented as a combination of cyber system emulation, which allows interactivity, and physical system (i.e., power grid) simulation that are tightly coupled for consistent cyber and physical behaviours.
However, its design and implementation require intensive expertise and efforts in cyber and physical aspects of smart power systems as well as software/system engineering. 
While many industry players, including power grid operators, device vendors, research and education sectors are interested, availability of the smart grid cyber range is limited to a small number of research labs.
To address this challenge, we have developed a framework for modelling a smart grid cyber range using an XML-based language, called SG-ML, and for ``compiling'' the model into an operational cyber range with minimal engineering efforts. The modelling language includes standardized schema from IEC 61850 and IEC 61131, which allows industry players to utilize their existing configurations.
The SG-ML framework aims at making a smart grid cyber range available to broader user bases to facilitate cybersecurity R\&D and hands-on exercises.

\end{abstract}


\thispagestyle{plain}

\section{Introduction}
%\todo[inline]{TBD} 
Emerging cyber threats against smart grid systems is one of the biggest concerns for ensuring dependable and trustworthy operation of our nations' critical infrastructure. In the last decades, we have witnessed a number of real-world incidents, such as Stuxnet attack in 2010~\cite{stuxnet}, Ukraine power plant attacks in 2015 and 2016~\cite{case2016analysis}, hacking against the US utility's control room in 2018, Venezuela blackout caused by cyber attacks in 2019~\cite{Venezuela2019}, ransomware attack against K-Electric in Pakistan~\cite{k-electric}. In order to counter and respond to such incidents, it is imperative to intensively evaluate the impact of various attack vectors, robustness of the system against these vectors, and to design response and recovery procedures.  

While it would be ideal to conduct such exercises and evaluations in the real system environment, it is impossible to avoid any potential impact on the availability and stability of the power grid operation. An alternate solution is to develop an isolated testbed using the same hardware as the real system, e.g., Electric Power and Intelligent Control (EPIC) testbed~\cite{epic}. However, such a hardware-based approach has inherent limitation in initial and operational cost, reconfigurability, scalability, and accessibility. Setting up and maintaining such a testbed is rather costly and hence may not be feasible for most organizations. Moreover, the system configuration and topology are fixed and difficult to modify or extend.
%, and its scale is usually far from the scale of the real system. Furthermore, the accessibility is limited to the on-site usage. 
In addition to these challenges, even in such an isolated testbed, experiments with high risk are often not permitted. For example, it would never be allowed to do the experiments similar to Aurora generator test done by Idaho National Lab researchers to demonstrate that cyber-originated attacks could physically destroy generators~\cite{aurora}.

Due to these reasons, high-fidelity, virtual environment for conducting {\color{black} interactive} cybersecurity experiments and exercises, often called cyber range, has attracted interest from both industry and academia. Smart grid cyber range is a virtual system that emulates cyber and physical systems of a smart grid, and can interact with human users in a real-time manner for conducting various experiments. Numerous efforts have been made to develop cyber range for smart grid systems in the recent years. Numerous efforts have been made to develop a cyber range for smart grid systems in recent years. Some of the efforts, such as~\cite{SoftGrid}, have limitations in terms of fidelity while others are designed to emulate one specific system or designed in one-off, in-house manner~\cite{lowcost,5759169, 2019817,7116592,costCPS,mashima2020design,roomi2020false,epictwin}, resulting in very limited accessibility. %Therefore, customization, scale-up of the testbed are still non-trivial, and portability and accessibility of the testbeds are limited. 

%requires intensive efforts and domain knowledge in both cyber and physical sides of the system.

To address these challenges and facilitate industry players and academic researchers to have their own cyber range on premise,  
%
%to make the cyber range easily reconfigurable and extensible, as well as to make the cyber range more portable and reproducible, 
we have developed a framework for automated smart grid cyber range generation, called {\it SG-ML} (Smart Grid Modelling Language)~\cite{sgml-techreport}. SG-ML defines XML-based modelling language for defining configurations of smart grid cyber range and provides the toolchain, {\it SG-ML Processor}, for parsing the models and, like a compiler, instantiating a cyber range according to the configuration. This paper focuses on the description of the SG-ML Processor toolchain. SG-ML is defined by using standardized models, such as IEC 61850 SCL (System Configuration description Language)~\cite{mackiewicz2006overview, tc572010iec} that power grid operators already have. Thus users in the power grid industry can utilize their existing assets to generate a digital replica of their smart grid, which can be, for instance, utilized for a red-team exercise to identify vulnerabilities without affecting the production system. Moreover, the XML-based model can be shared and customized in the open-source community. As a result, even users in other sectors can develop their own cyber range or reproduce it with minimal power grid expertise and engineering efforts.
%
%Since the SG-ML model is expressed in XML, it is human/machine-friendly, and thus can be customized or edited by human users. Furthermore, the model can be reproduced easily based on the user needs
%the model can be shared with community easily so that the cyber range can be reproduced 
%without intensive efforts or domain knowledge.
\footnote{SG-ML framework is open-sourced, along with the specification document and demo videos, at \href{https://github.com/smartgridadsc/CyberRange}{https://github.com/smartgridadsc/CyberRange}}

Different types of smart grid security testbeds (physical, hybrid and virtual) are developed over the years for research, training and experiment applications~\cite{smadi2021comprehensive,fovino2010experimental,barnes2009national,mcdonald2010modeling,SoftGrid,epictwin,roomi2023analysis}. 
%Some of the notable efforts are:
%1) testbed to enhance security on supervisory control and data acquisition (SCADA) systems~\cite{fovino2010experimental}; 2) SCADA testbed by Idaho National Laboratory to identify vulnerabilities in SCADA~\cite{barnes2009national}; 3) virtual control system environment to conduct cyber attacks on large-scale infrastructure~\cite{mcdonald2010modeling}; 4) software based testbed to experiment malicious command injection in IEC 61850 compliant substation systems~\cite{SoftGrid}; 5) physical smart grid testbed, such as Electric Power and Intelligent Control (EPIC) testbed~\cite{epic}; 6) EPIC TWIN~\cite{epictwin}, a digital twin of the EPIC testbed~\cite{roomi2023analysis}. 
%
Although these testbeds have different advantages and are utilized for different applications, limitations are found in one or multiple of deployability, configurability, scalability, and reproducibility. In particular, none of the existing efforts considers automated generation of smart grid cyber range to minimize user's burden on designing and implementing the system. To our knowledge, the SG-ML framework we have developed is the first to largely automate the generation of smart grid cyber range based on user-defined models. 
%Moreover, use of open-source software made the smart grid cyber range accessible for broader user community with the minimal cost. 
%While we admit that the generated cyber range may have limitations compared to the existing work using high-end simulators (e.g., real-time and high-fidelity dynamics simulation of physical plants), the generated cyber range is still sufficient for meaningful cyber attack experiments and exercises as demonstrated in Section~\ref{sec:crgen} and the demo video.  

Automated generation of cyber range from user-defined, standard-based models benefit industry players and practitioners in multiple ways. First, by generating a replica of the smart grid infrastructure, it can be utilized to assess potential vulnerabilities through extensive red-team testing without affecting the real system. Besides cybersecurity, such a cyber range can be utilized to evaluate compatibility and correctness of the power grid configuration (e.g., consistency among IED's (intelligent electronic devices) protection functions and PLC (programmable logic controllers) logics). Cyber range can also be used to conduct hardware-in-the-loop testing of PLCs and IEDs before deployment. Last but not the least, cyber range can be valuable resource for the cybersecurity hands-on training and education for technicians.

%there are limitations such as scalability of the system, flexibility in conducting experiments are limited. In these regards, the cyber range developed using SG-ML framework addresses the scalability issue (users can simulate small-scale to large-scale systems) and provides flexiblity in conducting wide range of cyber attacks. In this paper, we have demonstrated the most vulnerable and impactful attacks such as Man-in-the-middle (MITM) and false command injection (FCI) to highlight the advantage of the cyber range.}
%}

\section{Smart Grid Cyber Range Architecture}\label{sec:cyberrange}
\begin{figure}[b]
    \vspace{-4mm}
\centering
    \includegraphics[width=0.53\textwidth]{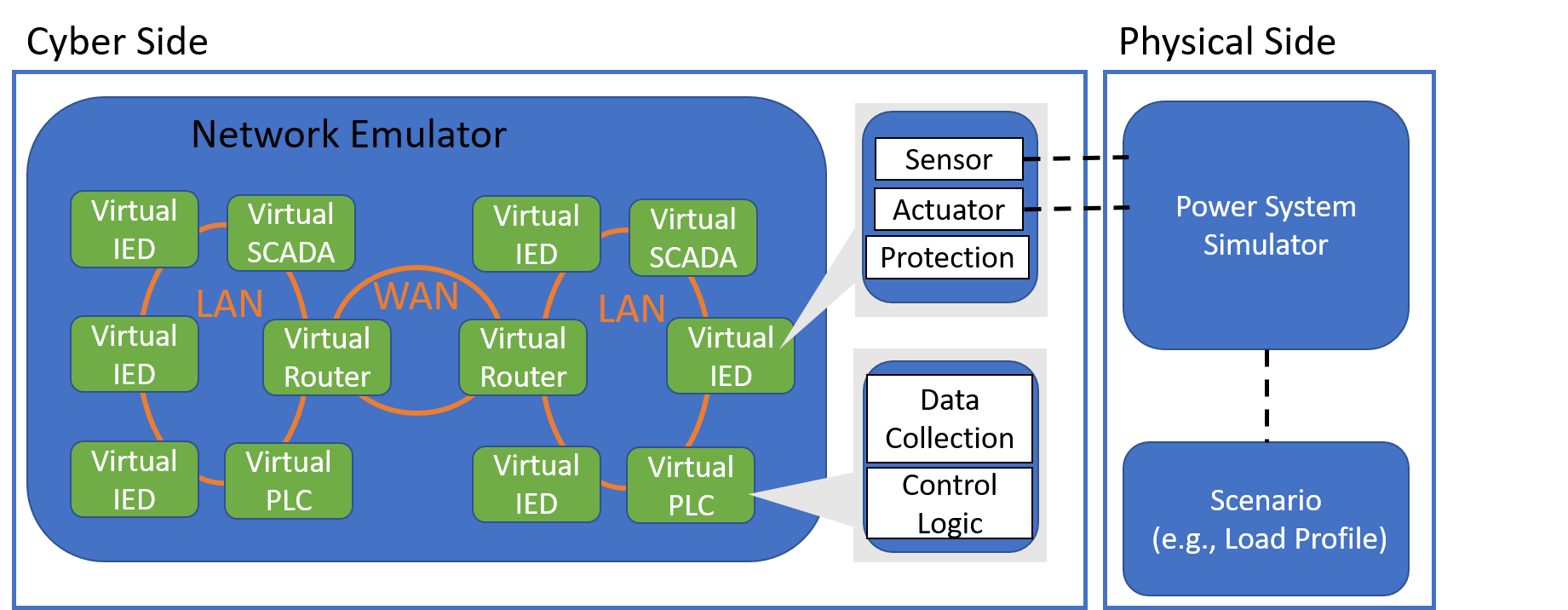}
    \caption{Typical Architecture of Smart Grid Cyber Range}
    \label{fig:cyberrange}
%    \todo[inline]{update figure that does not mention Mininet}
\end{figure}

We first summarize the components that are involved in a typical smart grid cyber range found in the literature, which is to be modelled and generated by the SG-ML framework. The high-level architecture is shown in Figure~\ref{fig:cyberrange}. 

Typically, cyber range for cyber-physical systems is implemented as {\color{black}a combination of cyber system emulation and a physical plant simulation}. In a smart grid cyber range for a flexible and interactive cyber attack experiments, the cyber side of the system should be implemented with a virtual network running a number of (virtual) smart grid devices, namely SCADA HMI {\color{black}(supervisory control and data acquisition human-machine interface)}, PLCs {\color{black}}, and IEDs {\color{black}}.
%so that they can interact with the user of the cyber range.
SCADA HMI is a user interface for human operators to see the status of the power grid system and, when necessary, sends control commands to the plant. IEDs work as sensors and actuators to collect power grid measurements (e.g., power, voltage, frequency, etc.) and operate on the physical devices, such as circuit breakers (CBs) and transformers. IEDs also implement protection function (e.g., over current / voltage protection) to protect the physical equipment from damage. IEDs communicate with other IEDs, PLCs, and/or SCADA HMI, using standard communication protocols, such as IEC 61850~\cite{mackiewicz2006overview}. PLCs are often utilized in smart grid system to implement automated control. PLCs collect measurements from one or multiple IEDs and then execute pre-configured control logic based on those inputs and sends control commands to IEDs. Control logic for the PLC is often programmed according to IEC 61131 standard~\cite{iec61131}.

The aforementioned cyber-side components interact with the power system simulator in real-time. For instance, when a virtual IED receives a control command to open a circuit breaker, it should take effect on the power flow status shortly after the receipt. {\color{black}The near-realtime interface between the cyber side and physical side can be implemented in multiple ways. Some commercial power system simulators have interactive API, and thus it can be utilized when the emulated virtual IEDs receives control commands. Alternatively, database or networking can be utilized. A cyber range discussed in~\cite{roomi2020false} utilizes database as the bi-directional (read/write) interface, and \cite{epictwin} utilizes publisher-subscriber communication to interact with the power system simulator. As the cyber range is mainly intended for interactive cyber attack exercise and experiments, all of these options are regarded sufficient in practice.} Besides, power system simulation models can be configured according to some scenarios, such as contingency, disruptions, as well as abnormal load profiles. 

%In this paper, focusing on smart power grid systems of the discussed architecture, we present the implementation of the tool that is used to generate such a cyber range based on user-defined models. 
%\todo[inline]{some related work discussion here}

%\todo[inline]{add description of interface between cyber and physical sides. also discuss real-timeness requirement}

%\todo[inline]{Add cyber range overview figure. then discuss components to be included in the cyber range}

\section{Automated Smart Grid Cyber Range Generation Framework}\label{sec:framework}
{\color{black}
\begin{figure}[h]
    %\vspace{-2mm}

\centering
    \includegraphics[width=0.47\textwidth]{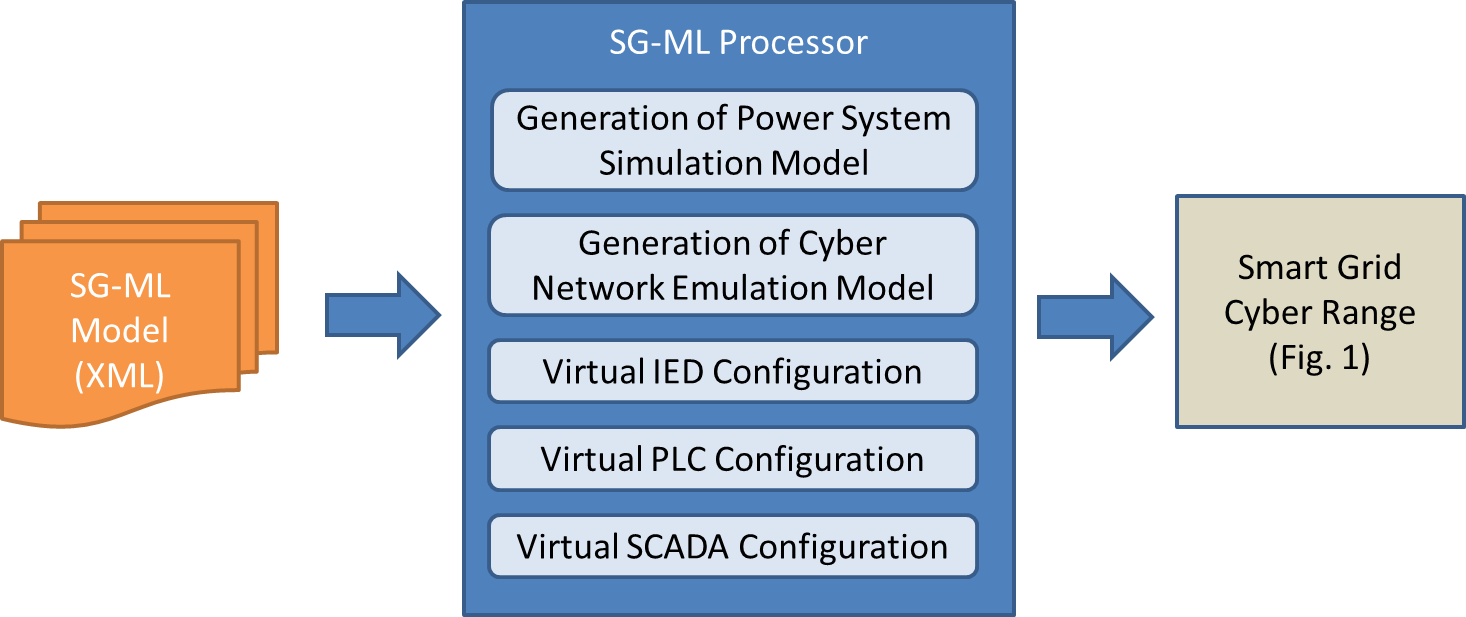}
    \vspace{-1mm}

    \caption{Overview of SG-ML Framework}
    \label{fig:sgml-overview}
    \vspace{-2mm}
%    \todo[inline]{replace this figure with a simpler version?}
\end{figure}
%In this section we discuss the overview of the framework for automated cyber range generation [anonymized]~\cite{sgml-techreport}.
The automated smart grid cyber range generation framework utilizes the XML-based modelling schema proposed by us, called SG-ML~\cite{sgml-techreport}. SG-ML is used to define cyber and physical configuration of smart grid cyber range, and the set of XML files are used as the input for the SG-ML Processor. SG-ML Processor processes the model files and generate an operational cyber range (Figure~\ref{fig:sgml-overview}). %This tool paper focuses on the description and demonstration of SG-ML Processor highlighted with blue color in Figure~\ref{fig:sgml-overview}
%
%In Figure~\ref{fig:sgml-overview}, the bottom section shows components consisting of the modelling language called SG-ML, the middle section corresponds to toolchain for automatically instantiating the smart grid cyber range according to the input SG-ML files, and the top section is the generated cyber range with the architecture illustrated in Section~\ref{sec:cyberrange}. This paper focuses on the introduction of SG-ML framework and toolchain. Further technical details can be found in another paper currently under review, thus left outside of the scope of this paper.
}
\subsection{SG-ML Specification}
%{\color{black}While detailed design and schema are explained in \cite{sgml-techreport}, a brief overview of SG-ML is provided for describing the toolchain.}
%
SG-ML is the modelling language for configuring smart grid cyber range and is defined as a set of XML schemas. Thus, the SG-ML model is both human and machine friendly. In other words, SG-ML models can be processed by software while human users can also write and modify them. 
%As illustrated in Figure~\ref{fig:sgml-overview}, 
We have decided to utilize standardized models used in the power grid sector, namely IEC 61850 SCL (System Configuration description Language) and IEC 61131-3 PLCopen XML, as well as the supplementary XML schema defined by us.    
IEC 61850 SCL schema is used in different kinds of configuration files for the IEC 61850 compliant substation systems (Table~\ref{tab:scl}).
This design choice allows power grid operators to recycle their own IEC 61850 SCL files corresponding to their own systems to generate a virtual replica for experiments and exercises. While we admit that preparing IEC 61850 SCL files from scratch is not trivial for users from other domains, this can be overcome by sharing a wide range of examples, including ones contributed by power grid experts, in a public repository. Since SCL files are well-structured XML files and well documented, customization of such template is not difficult. Moreover, such templates allows users to reproduce the same virtual environment for experiments, benchmarking and so forth.

 \begin{table}[t]
 \centering
 \renewcommand{\arraystretch}{1.1}
 \caption{{\color{black}Types of IEC 61850 SCL (System Configuration Description Language) Files and Description}}\label{tab:scl}
 \vspace{-2mm}
 \begin{footnotesize}
     \begin{tabular}{|p{2.2cm}|p{5.8cm}|}
         \hline
         \makecell{\textbf{Type of SCL File}} & \makecell{\textbf{Description}}\\
         \hline
                  \hline
SSD (System Specification Description) & Contains the overview of the substation structure as a single line diagram, voltage levels, bay levels, and functions of the substation.\\
         \hline
         SCD (System Configuration Description) & Contains complete description of the process including configuration for all IEDs present in a substation, structure or layout of the substation and a communication configuration section.\\
         \hline
         ICD (IED Capability Description) & Contains an overview of functionalities and engineering capabilities of an IED. It also contains the logical nodes (LNs) and the corresponding data types associated with the IED capability.\\
         \hline
         SED (System Exchange Description) & Contains information about the electrical connection between the two substations and the communication network information,  and semantics of IEDs involved in inter-substation communication.\\
         \hline
         
     \end{tabular}
     \end{footnotesize}
     \vspace{-3mm}
 \end{table}

%\todo[inline]{@suhail table for SSD, SCD, CID, SED etc. (by Nov 29)}

Based on our study, most of the static configuration of the smart grid can be derived from SCL files. For instance, an SSD file can be used to generate power flow simulation model while an SCD file can be used for cyber network topology to be emulated. An ICD file contains necessary information to define functionality of virtual smart grid devices, namely IEDs. An SED file defines cyber and physical connectivity between substations, and thus can be used to generate multi-substation topology. % along with SSD and SCD files. 
{\color{black}SG-ML framework can, in an unified way, support modelling and instantiation of various smart grid systems that differ in terms of the number of substations and the number of devices and topology involved in each substation.}  

\begin{comment}
The mapping of SCL files to the cyber range is summarized in Figure~\ref{fig:scl-cr}.

\begin{figure}[ht]
\centering
    \includegraphics[width=0.49\textwidth]{scl-cr.png}
    \caption{Mapping between IEC 61850 SCL to Smart Grid Cyber Range}
    \label{fig:scl-cr}
    \todo[inline]{@suhail, could you update this figure according to the namings used in paper? (by December 2)}
\end{figure}
\end{comment}

PLCs are the crucial component for automated control in industrial control systems, and they are often used in smart power grid systems. PLCs read measurements from power grid and then execute pre-configured control logic to initiate actuation. {\color{black}Such configuration are defined in IEC 61131-3 PLCopen XML~\cite{iec61131}, which expresses the control logic and variable definitions. }

{\color{black}
While the standardized models provide us with useful information for characterizing a cyber range, they are not sufficient. For instance, dynamic behaviour of the system, e.g., load profile and disturbance scenarios, cannot be configured in the SCL files. Thus, we have defined supplementary XML schema ({\it Power System Extra Config XML)}. Besides, parameters for IEDs' protection functions (described in Table~\ref{tab:protection}), such as alarm and trip thresholds, and the mapping between the cyber-side devices and physical-side device or information (e.g., which IED is measuring or controlling which transmission lines) are not included in the SCL files. Thus, we defined {\it IED Config XML} to incorporate the missing parameters. In addition, data sources and data points for SCADA HMI are not part of the SCL files. Hence, these can be defined in another supplementary XML schema {\it SCADA Config XML}. 
%s ({\it Power System Config XML}, {\it IED Config XML}, and {\it SCADA Config XML} in Figure~\ref{fig:sgml-overview}). Usage of these files is also elaborated later. 
%Details of SG-ML can be found in our online technical report~\cite{sgml-techreport}. 
In order to ensure user-friendliness, supplementary XML schemas are defined in a simple, straightforward format~\cite{sgml-techreport}.
Usage of the supplementary XML files will be illustrated in Figure~\ref{fig:processor-flow}.

}

\subsection{SG-ML Processor}
SG-ML Processor is a toolchain to parse SG-ML files that are used to generate cyber and physical model to be emulated in the cyber range. The components of the SG-ML Processor is illustrated in Figure~\ref{fig:sgml-overview} (the middle section). %{\color{black}This sections focuses on the high-level functionality of the modules in our toolchain. %, while leaving the discussion of the detailed processing logic and mechanisms to~[anonymized(under review)].} %The usage will be presented in the next section.

 \begin{table*}[t]
%\todo[inline]{@roomi for the left most column, could we add definition of acronym? is there any?}

 \centering
 \renewcommand{\arraystretch}{1.1}
 \caption{Protection Functions on Virtual IED}\label{tab:protection}
 \begin{footnotesize}
     \begin{tabular}{|p{3cm}|p{9cm}|p{5cm}|}
     \hline
         \textbf{\makecell{{\color{black}IEC 61850 Standard }\\{\color{black}Logical Node Name}}} & \textbf{\makecell{Description}}& %\textbf{\makecell{Thresholds in Supplementary Config File}}\\
         \textbf{\makecell{Thresholds in IED Config XML File}}\\
         \hline
         \hline

         PTOC (Time Over-current protection) & \underline {} Opens a circuit breaker when the amount of power flow exceeds the threshold. & Threshold limit for current, generally 3 to 4 times the nominal current \\
         PTOV (Over-voltage protection) & Opens a circuit breaker when the voltage on a bus exceeds the threshold & Threshold limit on bus voltage \\
         PTUV (Under-voltage protection) & Opens a circuit breaker when the voltage on a bus goes below the threshold & Lower limit of bus voltage  \\
         %PDOP & Reverse Power & Trip Threshold & Any small amount (trip) \\
         PDIF (Differential Protection) & Opens a circuit breaker when the current measurements at the 2 connected substations are different beyond the threshold & Threshold limit for differential current between two substations  \\
         CILO (Interlocking) & Prevents a circuit breaker to be closed when a certain circuit breaker is open & -  \\
         \hline
     \end{tabular}
     \end{footnotesize}
     \label{psf}
     %\vspace{-6mm}
     %\todo[inline]{@suhail,@roomi please correct or add. (by December 2)}
     \vspace{-4mm}

 \end{table*}

\noindent {\bf Generation of Power System Simulation Model:} 
An IEC 61850 SSD file contains information such as a single-line diagram of the power grid topology in a substation. Thus, SG-ML parses the SSD file and then generates a power system simulation model. In the current version, SG-ML Processor generates a simulation model for Pandapower~\cite{pandapower}, an open-source power system simulator. 
%{\color{blue} Since Pandapower is a steady-state simulator, which provides a snapshot of power grid status, the generated cyber range runs the simulaton in an iterative way with the updated parameters and topology (e.g., every 100ms - 1s depending on the system specification). Since the cyber range is intended for interactive experiments, the real-timeness of this level usually suffices.} 
%{\color{black} Since Pandapower is a steady-state simulator, which provides a snapshot of power grid status, the generated cyber range runs the simulation in an iterative way with the updated parameters and topology. In other words, any change of power grid status (e.g., loads) and control on power grid components takes effect on the next simulation cycle.} 
While an SSD file only contains topology in a single substation, it is often desired to simulate a larger system model consisting of multiple-substations. In such a case, IEC 61850 SED files, along with SSD files of the substations, are used. Typically, an SED file contains connectivity between a pair of substations. Our toolchain first combines multiple SSD files into a consolidated SSD file based on the connectivity derived from SED files. Then the consolidated SSD file is processed using the same tool to generate a multi-substation power grid physical model {\color{black}(see also Table~\ref{tab:scl})}.  
In order to run the power system simulation for the cyber range, we further need to provide information such as load profile and simulation scenarios. Simulation scenarios include disturbance and contingency, such as generator loss, line loss, etc. In order to incorporate these, the Power System Config Extra XML file is used. The XML file specifies the amount of load and circuit breaker status in a time series for each component in the simulation model. The power system simulator in the cyber range reads these parameters at each step of the simulation.

\noindent {\bf Generation of Cyber Network Emulation Model:} 
For each substation, cyber network model can be derived from IEC 61850 SCD file. An SCD file contains network addresses (including IP address and MAC address) of nodes, and connectivity between nodes (e.g., node-switch, switch-switch connections, etc.). These parameters are used to configure the network emulator. In the current version, we use Mininet~\cite{mininet} for the cyber network emulation. 
Similar to SSD files, each SCD file contains information about a single substation. Thus, to produce multi-substation cyber network model, we need to combine multiple SCD files. Typically, substations are connected through wide area network (WAN). %, which can be implemented as fibre optic cables or cellular network. 
The toolchain of the current version simplifies the emulation of WAN, and it is abstracted as a single switch connected to all substations.

\noindent {\bf Virtual IED Configuration:}
Virtual IEDs are implemented as an application written in C language using libiec61850 library~\cite{libiec61850}. A virtual IED implements communication using IEC 61850 protocols, including {\color{black}MMS (Manufacturing Message Specification), GOOSE (Generic Object Oriented Substation Event), R-GOOSE (Routable GOOSE) and R-SV (Routable Sampled Value). IEC 61850 MMS is utilized for communication between SCADA HMI and IEDs and PLCs and IEDs for conveying interrogation and control commands. GOOSE and R-GOOSE are utilized among IEDs to exchange device status information, while R-SV is used for sending power grid measurements among IEDs.} Virtual IEDs also implements popular protection functions, which are listed in Table~\ref{tab:protection}. Each virtual IED is instantiated by an IEC 61850 ICD file by enabling features defined in it. For instance, if the ICD file contains definition of logical node PTOV, over-voltage protection function is enabled. Moreover, if inter-substation protection function (logical node CILO) is defined in the ICD file, communication module for R-GOOSE and R-SV protocols are enabled. However, an ICD file alone is not sufficient because actual threshold for each protection function is not specified in the ICD file. Thus, to provide supplementary information, we have introduced IED Config XML. 
Virtual IEDs are connected to the power system simulator through {\color{black}an open-sourced} MySQL database. {\color{black} This works as a ``cache'' storing a set of key-value pairs,} for reading power grid measurements (voltages, power flow, etc.) and executing control (e.g., opening/closing circuit breakers). As it is necessary for each virtual IED to know the mapping between the naming of data item in the ICD file and the power system simulation output, such information is also part of self-defined IED Config XML~\cite{sgml-techreport}.

\begin{figure*}[!h]
\begin{tabular}{cc}

\begin{minipage}[c]{0.6\textwidth}

\centering
    \includegraphics[width=\textwidth]{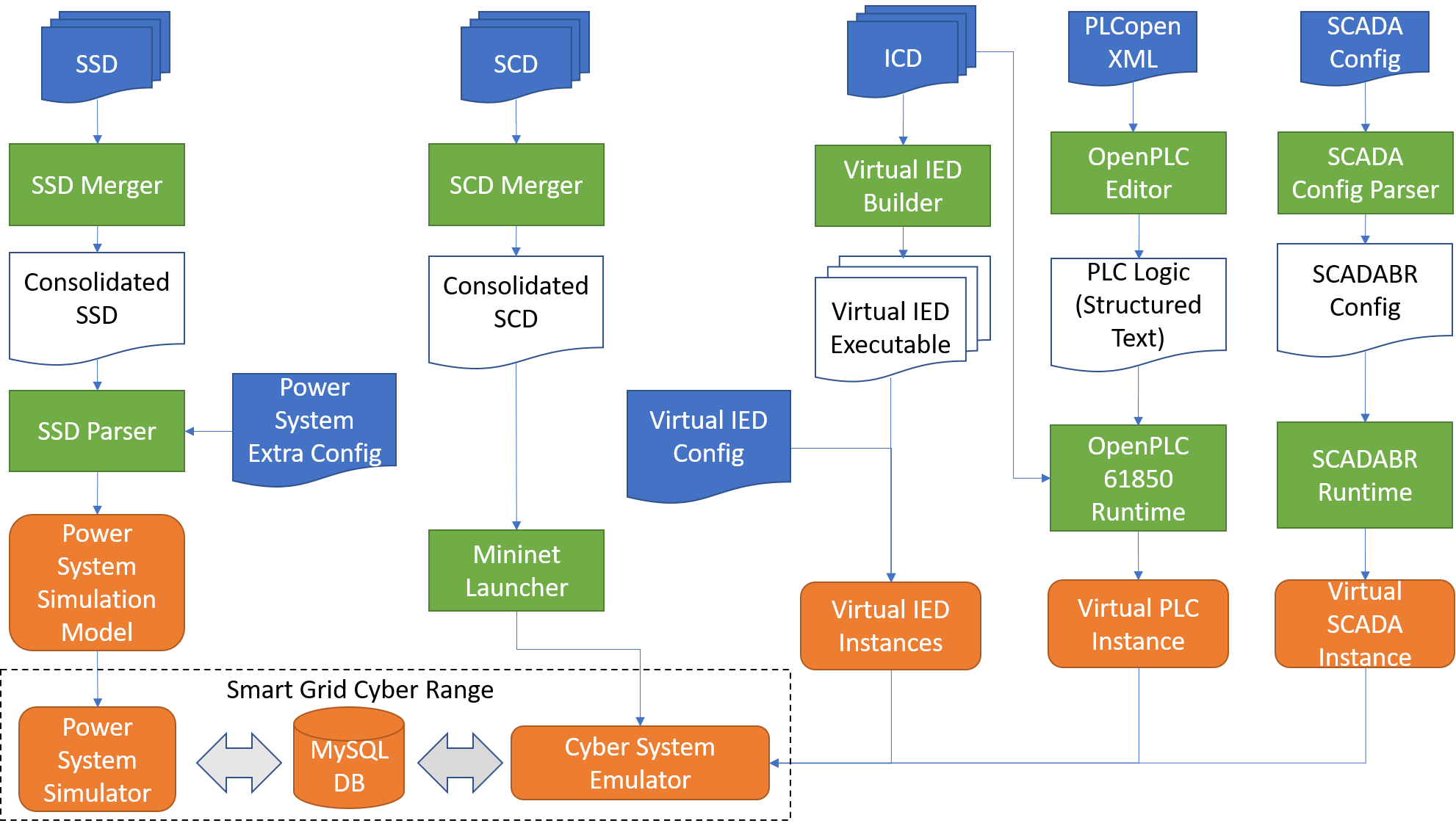}
    %\caption{SG-ML Processor Toolchain Flowchart. Gray, dotted boundary indicates the preparation phase before launching cyber range.}
    \label{fig:processor-flow}
\end{minipage}

\begin{minipage}[c]{0.4\textwidth}
\scriptsize
%\begin{table}[]
\centering
\renewcommand{\arraystretch}{1.1}
%\caption{System Detail and Required Software/Libraries}\label{tab:sysreq}
{\color{black}
\begin{tabular}{|p{2cm}|p{4cm}|}
\hline
\textbf{Module Name}                 & \textbf{Description} \\ \hline
\hline
SSD Merger & For multi-substation model, combine multiple SSD files into a consolidated SSD file\\
\hline
SCD Merger & For multi-substation model, combine multiple SCD files into a consolidated SCD file \\
\hline
SSD Parser & Generate power system simulation model from SSD \\
\hline
Mininet Launcher & Generate cyber topology from SCD and start virtual network \\
\hline
Virtual IED Builder &  Configure and compile virtual IED instance based on ICD\\
\hline
OpenPLC Editor / OpenPLC61850 &  Open-source PLC runtime and configuration tool~\cite{openplc-editor,openplc61850}\\
\hline
SCADA Config Parser &  Parse SCADA Config XML to generate JSON that can be imported to SCADABR\\
\hline
SCADABR & Open-source SCADA HMI for monitoring power grid status~\cite{scadabr}\\
 \hline
\end{tabular}
}
%\vspace{-4mm}
%\end{table}

\end{minipage}

\end{tabular}
    \caption{SG-ML Processor Toolchain Flowchart and Module Description}
    \label{fig:processor-flow}
    \vspace{-4mm}
\end{figure*}

\noindent {\bf Virtual PLC Configuration:}
%
%A PLC is an ICS device for automated control. 
Our cyber range uses an open-source PLC software, namely OpenPLC61850~\cite{softwarex}, for emulating a PLC. OpenPLC61850 supports Modbus communication protocol (for interacting with SCADA) and IEC 61850 MMS protocol towards IEDs. OpenPLC61850 requires a set of ICD files corresponding to the IEDs that it interacts with, as well as an IEC 61131-3 PLCopen XML file that contains control logic. %The PLCopen XML file can be translated into Structured Text file by using an open-source software like OpenPLC Editor~\cite{openplc}.
According to the configuration found in the SCD file explained earlier, OpenPLC61850 is started on nodes that run PLCs. %Then, from the OpenPLC61850's web interface, a user can upload the Structured Text file to the PLC. 

\noindent {\bf Virtual SCADA Configuration:}
A SCADA system offers an user-interface for a human user to monitor the system status and trigger manual control on a physical plant. Our cyber range utilizes an open-source software, called SCADABR~\cite{scadabr}. The settings on data source (e.g., PLCs) and data points has to be configured in SCADABR according to the user-defined model. We have implemented a script to translate the SCADA Config XML into a JSON format that SCADABR can import. 

{\color{black}The overall procedure of SG-ML Processor and the placement of each module (green rectangular) are found in Figure~\ref{fig:processor-flow}.}

\subsection{Limitation}\label{sec:limitation}
%\section{Limitations}\label{sec:limitation}
A cyber range generated by SG-ML relies on open-source tools for the sake of broader accessibility. %This decision poses some technical limitations. 
%
%One of the limitations to be noted is 
One limitation caused by this decision is that the power system simulation is not completely real-time or capable of simulating dynamics. {\color{black}Pandapower~\cite{pandapower} simulator we employed is a steady-state power flow simulation software, and it is a one-time solver that provides a snapshot of power grid status. Therefore, our cyber range runs it periodically (e.g., every 100ms) with the updated configuration and load profile. This implies that change of power grid status happens with this time granularity in a discrete manner. On the other hand, cyber range is intended for human-interactive cyber attack exercise and experiments, and SCADA HMI and PLCs are collecting data usually with second-level granularity. Thus, the time granularity and real-timeness of this degree are still acceptable in practice.} 
%Another drawback is that Pandapower cannot perform power system dynamics simulation (or also called transient-state simulation). This challenge could be addressed by incorporating commercial simulator, such as PowerWorld~\cite{powerworld}, Matlab SimuLink, or real-time simulator hardware, such as RTDS and OPAL-RT. {\color{blue}These enhancements, however, come with a drawback in terms of significant increase in the cost and the limitation in scalability.

%Another limitation is that the virtual IEDs and PLCs, while they emulate the functionality that are generally found in real devices, they don't run the firmware image of the real devices. Thus, attack vectors targeting  
%}   

%Another limitation on the cyber side implementation is reliance on Mininet. While Mininet is a convenient tool for emulating a network on a single PC, it may not be a best solution not only for scalability, which emulates a larger scale cyber range that spans over multiple physical nodes, but also flexibility to do hardware-in-the-loop experiment. For improving these aspects, we plan to explore Docker and its overlay network technologies.

%Last but not the lest, limitation is that the communication protocol supported is limited. In the substation network and inter-substation communication, only IE 61850 protocols are supported. On the other hand, SCADA HMI is limited to Modbus TCP. We plan to add more communication protocols, such as DNP3 and IEC 60870-5-104, in our future release.

\section{Demonstration of SG-ML and Cyber Range}\label{sec:crgen}

\subsection{Generation of EPIC Testbed Cyber Range}\label{sec:1-sub}
%Let us start with the most basic case with a single substation. 
%\todo[inline]{@roomi let us add a bit of more detail about EPIC.}
{\color{black}EPIC~\cite{epic,epic-cpss} is a state-of-the-art smart grid security testbed hosted by Singapore University of Technology and Design. This testbed is utilized for research and training to design a safe and secure critical infrastructure. 
%Furthermore, this testbed provides flexibility to conduct cyber security experiments as a measure to evaluate the defense mechanisms. 
Therefore, we consider EPIC testbed for our demonstration.}
%Let us start with the case with EPIC testbed, whose 
%
{\color{black}The} topology and configuration are available at~\cite{epic,epic-cpss}. EPIC testbed consists of 4 segments, namely generation, transmission, microgrid, and smart homes.
{\color{black}The power generation is performed through two generators (in generation segment) and PV and batteries (in microgrid segment). This combined generation reflects the modern-day power grid with conventional power sources and renewable energy sources (RES). The smart homes consist of controllable loads.} Each of these segments include multiple IEDs connected the power grid devices and is monitored by a SCADA HMI. While the original EPIC testbed includes multiple PLCs, in the cyber range we consider one PLC that mediates communication between SCADA HMI and IEDs (called CPLC). 
%Another simplification is that, while 
In the real power grid, these segments belong to different substations. However, 
%because of the small scale of the system and 
following the EPIC testbed setting, we consider all belong to a single substation.

%Here, we consider the typical configuration of a distribution level substation, whose physical topology is illustrated in~\cite{partha2019}.

\begin{comment}
For this model, SG-ML model files used as the input include the following:
%\todo[inline]{@bennet update the number of files  in EPIC model (by nov 29) - done.}
\begin{enumerate}
    \item 1 IEC 61850 SSD file
    \item 1 IEC 61850 SCD file
    \item 9 IEC 61850 ICD files
    \item 1 PLCopen XML file
    \item 1 SCADA Config XML file
    %\item 1 supplementary config file for power system simulator
    %\item 1 supplementary config file for IED parameter
    %\item 1 supplementary config file for cyber-physical mapping
    \item 1 Power System Extra Config XML file
    \item 1 IED Config XML file (thresholds and cyber-physical mappings)
\end{enumerate}
\end{comment}

%In order to generate the SCL files, we utilized a commercial software called SCL Matrix~\cite{SCLmatrix}, but 
The SCL 
files can be generated with any tool or can be obtained from the real system or an open-source community. %Power grid operator could utilize the existing SCL files. 
%The overall flow is found in Figure~\ref{fig:processor-flow}. 
%
\begin{figure*}[th]
% \begin{tabular*}{cc}

  \begin{minipage}[b]{0.55\textwidth}
   \centering
    \includegraphics[width=0.7\linewidth]{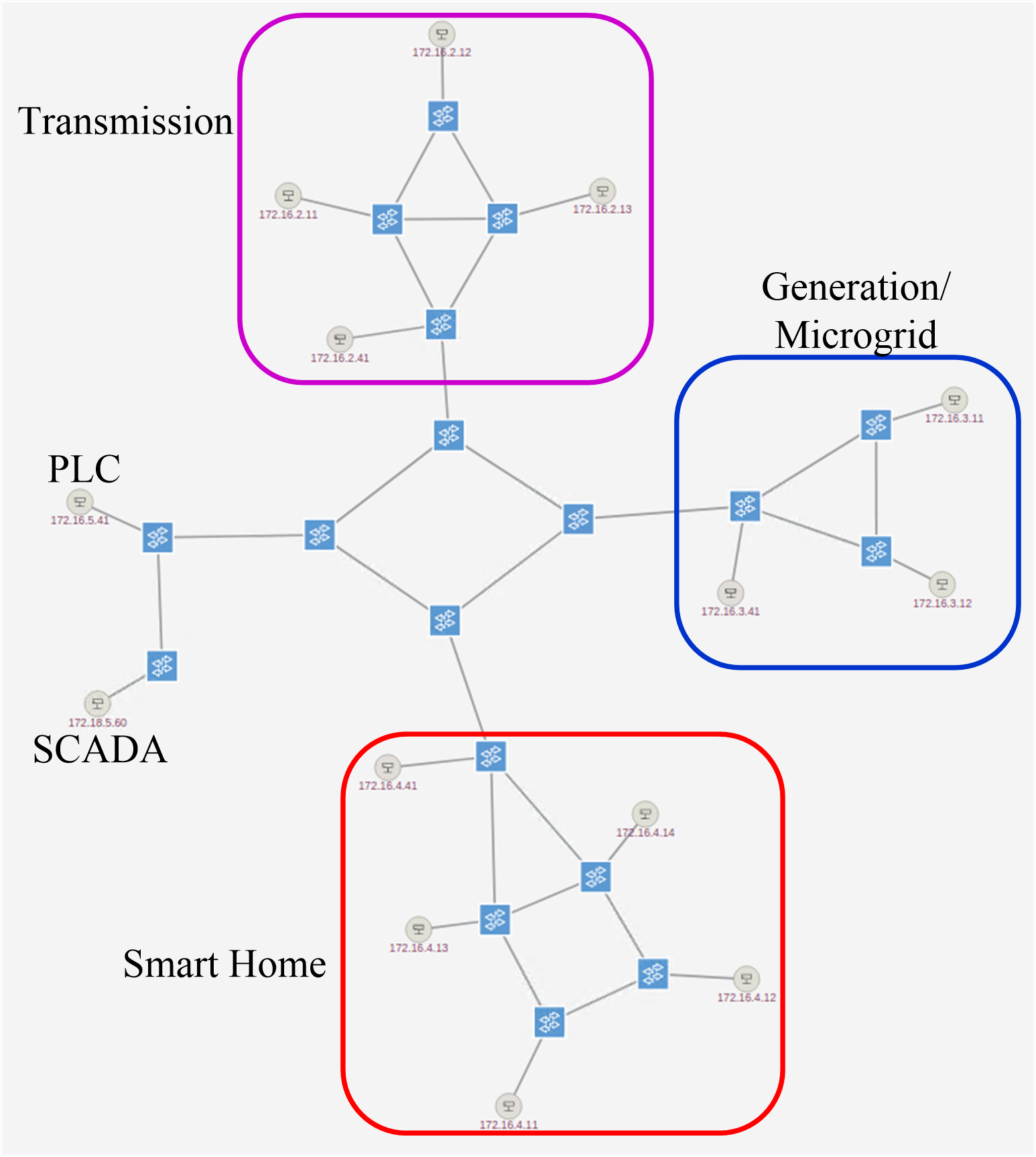}
    \caption{Generated Cyber Network Topology on Mininet (EPIC Model) generated using ONOS~\cite{onos}. Rounded rectangles show mapping to the EPIC testbed.}
    \label{fig:1sub-mininet}
   \end{minipage}
   \hfill
   \begin{minipage}[b]{0.45\textwidth}
   \centering
    \includegraphics[width=\linewidth, height=\linewidth]{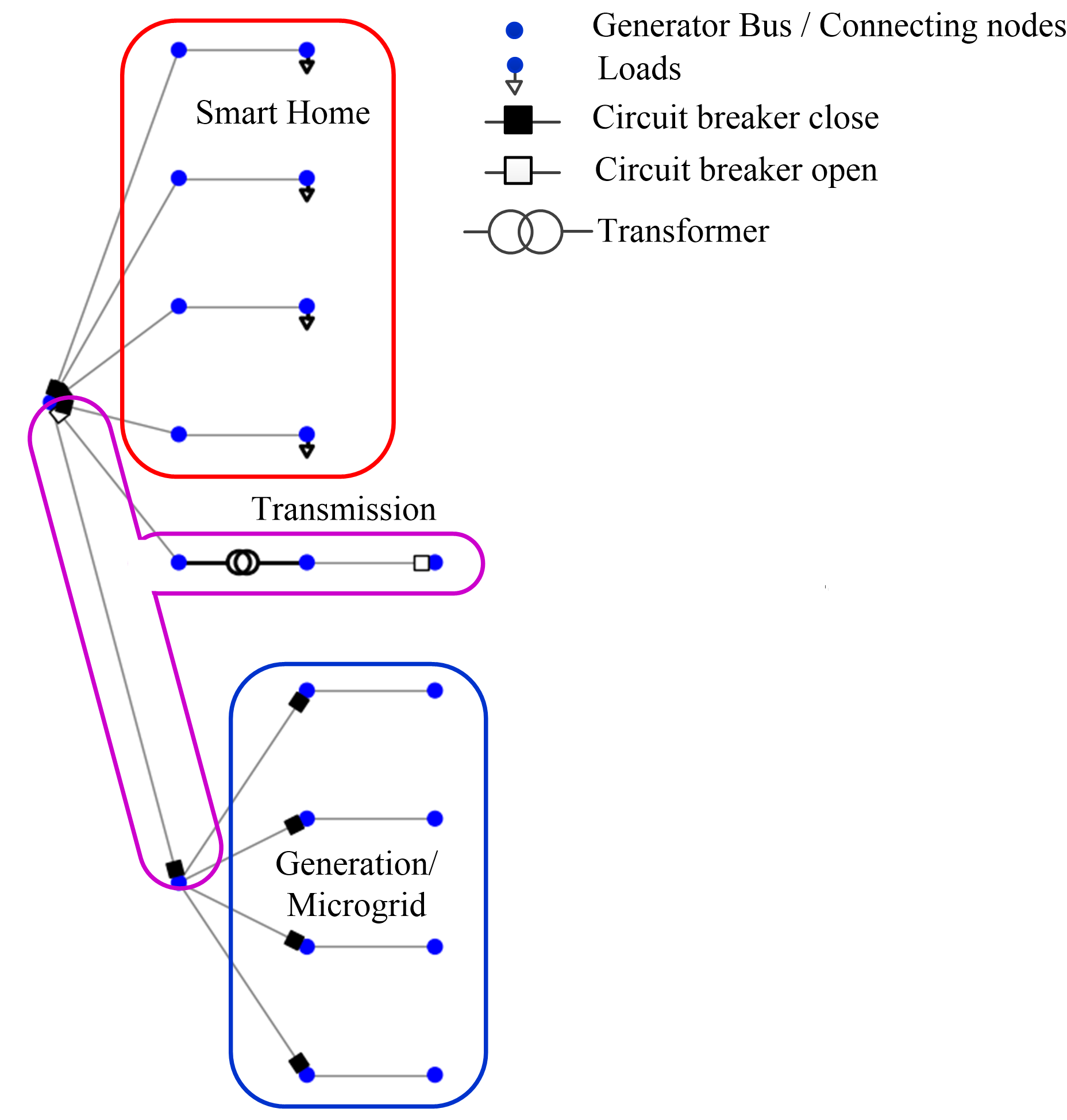}
    \caption{Generated Power System Topology on Pandapower (EPIC Model). Rounded rectangles show mapping to the EPIC testbed.}
    \label{fig:1sub-pp}
   \end{minipage}
% \end{figure}

% \end{tabular*}
\vspace{-4mm}
\end{figure*}
%
%
%Before starting the cyber range, we need to conduct the following tasks. 
Our toolchain conducts the preparation tasks in a largely automated manner, such as generation of the power system simulation model, cyber network emulation model, and/or  virtual IED, PLC, and SCADA HMI instances
(Figure~\ref{fig:processor-flow}).
%See the procedures in the grey, dotted rectangular in
%Figure~\ref{fig:processor-flow}). 
%Some of these tasks, such as generation of consolidated SSD and SCD, are not necessary when we construct a single substation cyber range. Moreover, if the cyber range does not include PLC or SCADA HMI, corresponding tasks are not necessary. %Virtual IED executables are generated based on the input ICD files. Using IEC 61850 ICD file, information model and dataset definition that each IED manages are configured, and then, virtual IED executables are compiled accordingly.     
%\todo[inline]{discuss preparation tasks before starting mininet here.}

%Once the preparation phase is done, we can instantiate and launch the cyber range. The next step is to start emulated substation network on the Mininet emulator~\cite{mininet}. 
The cyber topology of the EPIC testbed model is shown in Figure~\ref{fig:1sub-mininet}. The scripts in our toolchain parse an SCD file (consolidated SCD, in case of multi-substation model) and then extract necessary information into an intermediate JSON file, which is then passed to the script to configure and start the Mininet emulator. The same script also launches virtual IED executables on the specified virtual nodes on the Mininet.

%\todo[inline]{@benn please update EPIC mininet topology figure}

%\begin{figure}[!h]
%\centering
%    \includegraphics[width=0.9\linewidth]{1-substation-mn.png}
%    \caption{Generated Cyber Network Topology on Mininet (EPIC Model)}
%    \label{fig:1sub-mininet}
%\end{figure}

%This step also instantiates virtual nodes to run virtual smart grid devices (i.e., IEDs, PLCs, and SCADA). Figure~\ref{fig:1sub-mn}.  

After the Mininet emulator is started, the virtual PLC and a SCADA HMI is started on the respective virtual node on Mininet. As open-source tools (OpenPLC61850~\cite{softwarex} and SCADABR\cite{scadabr}) are used, the starting process has to be done manually. The user can access the command-line terminal and/or web interface of the corresponding virtual node and then start the necessary program. %We note that, if PLC or SCADA is not involved in the topology, this step is optional. 
PLC logic in Structured Text format can be uploaded to the OpenPLC runtime and then started. The main responsibility of CPLC in the EPIC testbed is to mediate the communication between IEDs and SCADA HMI. Therefore, we also configured the virtual PLC accordingly. Note that OpenPLC61850 also requires ICD files of IEDs that the PLC interacts with. Regarding SCADABR, after starting up, the user can upload the SCADABR Config JSON data that defines data sources and data points. 

After confirming the communication among the virtual devices,
%on the emulated network, 
we can start the power flow simulator. We utilize Pandapower simulator, and the topology is automatically defined according to the SSD file. Our script generates the topology by parsing the SSD file, and then, by using the information from the configuration file, a sequence of power system simulation models is created (Figure~\ref{fig:1sub-pp}). 
The models in the sequence are executed at a designated time slot (e.g., 100ms interval). %The simulation model for this substation is found in Figure~\ref{fig:1sub-pp}.

%While we demonstrated a small scale model for the sake of illustration
Our toolchain can support smart grid system including multiple substations, beyond the scale of the EPIC model. Cyber and physical connectivity among substations and communication models are defined by using IEC 61850 SED files (see Table~\ref{tab:scl}). Based on our experiments, a commodity desktop PC with Intel Core i9 Processor and 16GB RAM can host a 5-substation model including 104 virtual IEDs with 100ms power flow simulation interval. It is also possible to define a cyber range spanning over multiple nodes to scale up further.

\subsection{Cyber Attack Case Studies}\label{sec:casestudy}

%{\color{black}
As the main purpose of a cyber range is to conduct interactive cyber attack experiments and exercises, in this section we discuss how the cyber attack case studies can be conducted on the cyber range. Owing to the limitation of space, we leave the demonstration to the demo video found at \url{https://github.com/smartgridadsc/CyberRange}.
%}
%
%present cyber attack case studies to show how the generated cyber range can be utilized. 
%\begin{comment}
 
Among a wide range of attack vectors, we focus on false command injection and man-in-the-middle attacks. The former can cause direct and immediate impact on power grid stability as demonstrated in the 2015 Ukraine incident~\cite{ukraine}, and the latter is a versatile building block for mounting a wide range of attacks, such as false data injection and alarm suppression. %However, we note that the generated cyber range can be used for other types of cyber-originated attacks, including denial-of-service attacks, replay attacks, etc.
We note that attacks that can be experimented are not limited to these two. Users can utilize any penetration testing tool like Nmap and Metasploit on a virtual node of the cyber range or on their own devices connected to the cyber range. 
%
%\end{comment}
%{\color{black}Demo videos showing cyber range generation and attack case studies are available on \url{https://osf.io/n6ycb/?view_only=f430f148f70c49b0860eb8ad9e5544ce}.}

%\subsection{False command injection (FCI)}
\noindent\textbf{False Command Injection (FCI) Attack: }
%Let us first consider an attacker who inject false command to the network by impersonating a legitimate SCADA HMI, like the 2015 Ukraine incident~\cite{ukraine} % or by compromising any node IED in the network and modify the values reported by the IED to PLC/SCADA HMI. %In this paper, the latter type of FCI attack is demonstrated.
%
Assuming that the attacker has compromised one of the nodes in the system and run malwares like CrashOverride~\cite{CrashOverride} to transmit fake IEC 61850 MMS commands. 
Attack of this sort can be experimented by running an IEC 61850 MMS client (e.g., IEC61850bean~\cite{iec61850bean}) on a node in the cyber range to emit standard-compliant command messages. 
%The attacker establishes a connection to a target IED using the IEC 61850 MMS client software. By identifying the controllable data attribute, the attacker may try to change the {\color{blue}circuit breaker (CB)} status (open/close). 
Once the IED receives a circuit breaker (CB) open command, for instance, the corresponding CB is operated, and the power flow change is calculated by the power flow simulator. % (Figure.~\ref{linedis}). 
%The de-energized line due to the malicious command injection and the modified CB status are displayed in Figure.~\ref{linedis}.

\begin{comment}
\begin{figure}[htpb]
\centerline
{\includegraphics[width=1\linewidth]{normala.jpg}}
\caption{Manipulation of Circuit Breaker Status Using IEC61850bean}
\label{fakecb}
\vspace{-1mm}
\end{figure}
\end{comment}

%\begin{figure}[htpb]
%\centerline
%{\includegraphics[width=\linewidth]{normalb.jpg}}
%\caption{Modified Line Status Seen on SCADA HMI}
%\label{linedis}
%\vspace{-1mm}
%\end{figure}

\begin{figure}[!h]
\centering

%        \vspace{-2mm}
\includegraphics[width=1\linewidth]{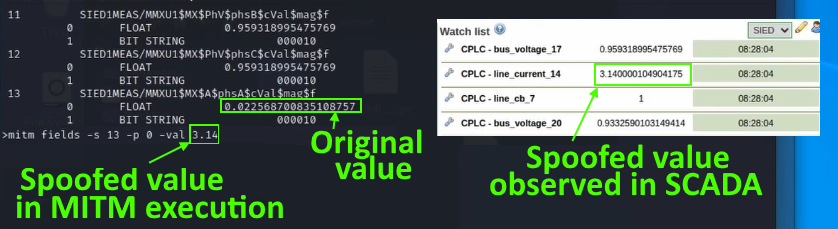}
    \caption{MITM Attack on a Power Grid Measurement}
    \label{fig:mitm}
    \vspace{-3mm}
\end{figure}

%\subsection{Man-in-the-middle (MITM) attack}
\noindent\textbf{Man-in-the-middle (MITM) Attack: }
%Next, we demonstrate man-in-the-middle attack. 
Typically man-in-the-middle (MITM) attack is mounted by using a strategy called ARP {\color{black}(Address Resolution Protocol)} spoofing. {\color{black}This confuses the mapping between a device's logical (IP) address and physical address}. Using ARP spoofing, an attacker can mislead the traffic to itself for interception and manipulation. %(Figure~\ref{fig:mitm}). 
%We implemented a script to mount such an MITM attack against IEC 61850 MMS protocol, which is shown in Figure~\ref{fig:mitm}. Using such a tool, we experimented a case where an MITM attacker modifies the payload of report messages sent by an IED to a PLC. 
As a consequence, the attacker could possibly mislead the SCADA HMI or the PLC to confuse the plant control (Figure~\ref{fig:mitm}).

%\todo[inline]{@benn let us add screenshots of MITM (by nov 29)}
\begin{comment}
\todo[inline]{ideally, show attack scenario that is not discussed in TII paper

- command injection (I copied one commented out in TII, sounds ok?)

- Some attack targeting IED protection logic?
}
\end{comment}

%\begin{comment}
%\end{comment}

%\todo[inline]{We could add some more later, if space allows. e.g., device is generic, not emulating specific model}

\section{Conclusions}\label{sec:conclusion}
We introduced a novel framework for modelling and automated generation of smart grid cyber range, called SG-ML. SG-ML framework makes a smart grid cyber range accessible to broader user base, including power grid industry, smart grid device vendors, education sectors, and academia. In particular, power grid operators can use our tool to generate a virtual replica of the real infrastructure to conduct intensive red-team testing, validation of configurations, compatibility testing, and so forth.
We have open-sourced the tool along with example models for getting real-world feedback for future enhancement. Enhancement of flexibility by using other virtualization technologies, such as Docker, is part of our future work. We also plan to develop a cloud-based cyber range service based on our  framework for enhanced accessibility and scalability.  

\section*{Acknowledgment}
This research is supported in part by the National Research Foundation, Singapore, Singapore University of Technology and Design under its National Satellite of Excellence in Design Science and Technology for Secure Critical Infrastructure Grant (NSoE\_DeST-SCI2019-0005), and in part 
by the National Research Foundation, Prime Minister’s Office, Singapore under its Campus for Research Excellence and Technological Enterprise (CREATE) programme.

\balance

\bibliographystyle{IEEEtran}
\bibliography{conference_101719}

\end{document}